# Polaronic Optical Transitions in Hematite (α-Fe$_2$O$_3$) Revealed by First-Principles Electron-Phonon Coupling

Jacob L. Shelton and Kathryn E. Knowles*

Department of Chemistry, University of Rochester, Rochester, NY 14627

*Corresponding Author e-mail: kknowles@ur.rochester.edu

## Abstract

Polaron formation following optical absorption is a key process that defines the photophysical properties of many semiconducting transition metal oxides, which comprise an important class of materials with potential optoelectronic and photocatalytic applications. In this work, we use hematite (α-Fe$_2$O$_3$) as a model transition metal oxide semiconductor to demonstrate the feasibility of direct optical population of band-edge polaronic states. We employ first-principles electron-phonon computations within the framework of the DFT+$U$+$J$ method to reveal the presence of these states within a thermal distribution of phonon displacements and model their evolution with temperature. Our computations reproduce the temperature dependence of the optical dielectric function of hematite with remarkable accuracy and indicate that the band-edge optical absorption and second-order resonance Raman spectra arise from polaronic optical transitions involving coupling to longitudinal optical phonons with energies greater than 50 meV. Additionally, we find that the resulting polaron comprises an electron localized to two adjacent Fe atoms with distortions that lie primarily along the coordinates of phonons with energies of 31 and 81 meV.

**Introduction**

Semiconducting transition metal oxides are a promising class of materials for the development of next-generation light-harvesting devices due to their exceptional photochemical stability and sustainable large-scale synthesis from earth-abundant precursors, and have thus been the subject of intense research throughout the past two decades.[1-4] However, the scope of solar energy conversion applications for transition metal oxides has been limited by their inherent tendency to form polarons, quasiparticles comprised of a charge carrier (electron or hole) bound to a proximal distortion of the host lattice.[5,6] Polarons arise from strong carrier-phonon interactions and are primarily classified by the spatial extent of their characteristic lattice distortion. A large polaron is one whose carrier wavefunction and lattice polarization extend beyond a single lattice constant. Conversely, a small polaron is contained within the volume of approximately one primitive unit of the crystal structure.[6] The size of a polaron influences its mobility: large polarons exhibit band-transport properties similar to free carriers, whereas the mobility of small polarons is limited to thermally-activated carrier-hopping processes. The theoretical description of polarons dates back to 1933, when Landau first proposed their existence;[7] however, atomically precise descriptions of polaronic states and their formation in transition metal oxides that fully account for extended solid-state structure and can accurately reproduce experimental spectroscopic data are lacking. Such models are crucial to developing a complete understanding of the electronic structure and photophysics of transition metal oxide semiconductors.

Transition metal oxides are particularly challenging to model within the context of band theory due to their unique electronic structure. In conventional crystalline semiconductors (e.g., II-VI, III-V, and group-14 elemental systems), the electronic structure of the band gap is dictated by hybridization of valence $s$ and $p$ orbitals of the atomic constituents of the crystal. Significant orbital mixing within the *sp* subspace results in dispersive bands that accommodate efficient covalent charge transport. In contrast, the electronic structure of transition metal oxides is strongly influenced by additional interactions involving the 3*d* orbitals of the transition metal ions, particularly those with an open-shell configuration. As a result, the near-gap bands are generally more ionic in character and, thus, weakly dispersive.[8,9] This phenomenon has a series of profound consequences: (i) weakly dispersive bands give rise to a high density of strongly localized electronic states within the lattice, (ii) charge carriers (electrons or holes) occupying these bands experience relatively high effective masses and strong on-site carrier-carrier interactions, and (iii) the energy and mobility of these carriers become inherently coupled to the motion of the surrounding nuclei (i.e., phonons).[9] Altogether, these consequences comprise two of the most pervasive and fundamental challenges to computational modeling of the electronic structure of transition metal oxides, namely strong carrier-carrier and carrier-phonon interactions.

Localized electronic states in solids, such as polarons, are notoriously difficult to describe within the confines of approximate density functional theory (DFT). This difficulty is due, in part, to the failure of standard exchange and correlation functionals to describe strong Coulombic interactions between localized electrons. Such self-interaction errors typically lead to an over-delocalization of valence electrons in the ground state predicted by standard DFT. Open-shell transition metal oxide semiconductors are particularly susceptible to these errors due to the strongly correlated nature of their metal *d*-orbital subspace. Fortunately, this problem is now routinely mitigated by the widely used DFT+$U$ and DFT+$U$+$J$ methods, whereby an on-site Coulombic repulsion (Hubbard) parameter $U$ and a site exchange (Hund's) parameter $J$ are added to the total energy functional of standard DFT. Given the additive nature of the corrective functionals, implementation of these methods into existing computational codes is simple;

however, the task of choosing appropriate values for the *U* and *J* parameters can prove challenging. Many authors choose to tune the parameters until a particular computed observable agrees with experimental data, while others simply opt to use previously reported values successfully applied in modeling similar systems.[10] Although prevalent in literature, both routes are poorly justified. Both *U* and *J* should be taken as inherent properties of the system itself and of the particular pseudopotentials and functionals employed in the model. A more rigorous approach is to explicitly calculate the parameters *ab-initio* by way of density functional perturbation theory (DFPT) in order to ensure the corrected functional precisely accounts for the magnitude of self-interaction errors present in the specific system of interest.[11-13]

Inclusion of vibrational degrees of freedom presents further challenges to accurate modeling of polaronic states using traditional computational treatments. Thermal lattice vibrations are described by a stochastic superposition of phonons and represent a departure from Bloch periodicity. The resulting disorder causes states of previously well-defined wavevector to become mixed, often leading to a significant renormalization of the electronic structure. Standard DFT is severely limited by the fundamental approximation that atomic nuclei are completely immobile in the crystal lattice (i.e., fixed at their equilibrium positions). As such, the effects of dynamic thermal fluctuations on the nuclear potential are neglected.[14] Among the most significant of these neglected phenomena is the contribution of phonon-coupled optical transitions to computed optical spectra. These effects are crucial to the accurate modeling of the temperature dependence of optical functions of materials, such as transition metal oxides, that exhibit strong carrier-phonon coupling and polaron formation in ground and/or photoexcited states.[14, 15]

Here, we use hematite ($\alpha$-$Fe_2O_3$) as a model transition metal oxide to develop a method to compute temperature-dependent electronic structures of strongly correlated semiconductors based on a first-principles atomically precise treatment of carrier-phonon coupling. Among transition metal oxides, hematite has emerged as one of the most well-studied due to its potential as a photoanode for water oxidation.[16-25] However, bulk charge transport in hematite is constrained by small polaron formation. Previous studies have suggested that both electrons[26-32] and holes[31-34] form polarons in hematite following excess charge injection or oxygen-vacancy doping. Furthermore, recent transient extreme-ultraviolet (XUV)[35-38] and pump-push photocurrent (PPPC)[39] spectroscopy studies have shown that small polarons form alongside thermal relaxation of photoexcited electrons. Our group previously demonstrated that small polaron bands can arise in a pristine hematite lattice without the need for prior charge injection, defect doping, or photoexcitation.[40] The computational methods developed here reveal several highly-localized optical transitions that appear near the band-edge of hematite upon inclusion of a thermal distribution of atomic displacements along phonon eigenvectors. By varying the temperature that governs the thermal distribution of phonon displacements in the lattice, we can reproduce accurately the thermal difference spectra of hematite we reported previously.[40] This work unambiguously confirms that these band tail transitions give rise to the formation of electron small polarons through strong coupling with longitudinal optical phonons.

**Theoretical Background**

*The DFT+U and DFT+U+J Framework*

The DFT+$U$ method is characterized by the addition of a parametrized correction to the total energy functional of standard DFT that is intended to mitigate self-interaction errors arising from strongly correlated electronic states. The resulting DFT+$U$ functional can be expressed as

$$E_{\text{DFT+U}} = E_{\text{DFT}}[\text{n}(\mathbf{r})] + E_{\text{U}}[\{\text{n}_{mm'}^{I\sigma}\}] \tag{1}$$

where $E_{\text{DFT}}$ is the standard DFT total energy functional, n(**r**) is the electron density, $E_{\text{U}}$ is the Hubbard correction, and n$_{mm'}^{I\sigma}$ are the orbital occupation matrix elements for each atom $I$ within the Hubbard manifold.[41-43] Here, $\sigma$ represents the electron spin index, ↑ or ↓. In the original formulation proposed by Anisimov and coworkers, $E_{\text{U}}$ was not rotationally invariant with respect to the localized Hubbard subspace and, as such, depended heavily on the choice of basis for the Hubbard orbital wavefunctions.[13] Today, the most widely used DFT+$U$ method is the rotationally-invariant form, first developed by Liechtenstein and coworkers[44, 45] and further simplified by Cococcioni and de Gironcoli.[11] In this formulation, $E_{\text{U}}$ is expressed as

$$E_{\text{U}} = \sum_{I,\sigma} \frac{U^I}{2} \text{Tr}[\mathbf{n}^{I\sigma}(1-\mathbf{n}^{I\sigma})] \qquad (2)$$

where $U^I$ is the atomic Coulomb repulsion and $\mathbf{n}^{I\sigma}$ is the occupation matrix of the Hubbard atom $I$. In this form, $E_{\text{U}}$ effectively acts as a penalty function that favors either full or null occupancy of the Hubbard subspace over fractional occupations, thus driving the system toward a state more akin to that predicted by exact DFT. The elements of $\mathbf{n}^{I\sigma}$ take the general form of a projection of the electronic Kohn-Sham valence wavefunctions ($\psi_{\mathbf{k}v}^{\sigma}$), denoted by crystal momentum **k** and band index $v$, onto a choice of localized basis functions.

$$\text{n}_{mm'}^{I\sigma} = \sum_{\mathbf{k},v} f_{\mathbf{k}v}^{\sigma} \langle \psi_{\mathbf{k}v}^{\sigma} | \hat{\text{P}}_{mm'}^{I} | \psi_{\mathbf{k}v}^{\sigma} \rangle \qquad (3)$$

In equation 3, $f_{\mathbf{k}v}^{\sigma}$ represents the Fermi-Dirac occupation number. Throughout this work, we employ a basis of atomic iron (Fe$^0$) $d$ orbitals, $\varphi_m^I$, where the index $I$ now specifically denotes iron atomic sites within the lattice. As such, the projection operator takes the simple form given in equation 4.

$$\hat{\text{P}}_{mm'}^{I} = |\varphi_m^I\rangle\langle\varphi_{m'}^I| \qquad (4)$$

In this work, we determine the value of $U^I$ for the iron $3d$ subspace of hematite within the framework of the linear response approach developed by Cococcioni and de Gironcoli.[11] The aim is to compute, from first principles, a value for the Hubbard parameter that corrects for the unphysical curvature of the total energy induced by non-integer occupation numbers. This goal is accomplished by way of constrained density functional calculations that provide a means of tracking the total energy of the system as a function of the local occupation of the Hubbard atoms. In practice, these constraints are implemented by applying localized perturbations, $\alpha^I$, to the Kohn-Sham potential of individual Hubbard sites, $I$, isolated within a supercell. Thus, the response function given in equation 5 affords a quantitative measurement of the total energy curvature with respect to changes in the on-site occupancy of the Hubbard manifold.

$$\chi_{IJ} = \frac{\partial^2 E}{\partial \alpha_I \partial \alpha_J} = \frac{\partial n_I}{\partial \alpha_J} \quad (5)$$

Elements of the response matrix are calculated as numerical derivatives by evaluating the occupations ($n^{I\sigma} = \text{Tr}[\mathbf{n}^{I\sigma}]$) subject to a discrete set of applied potentials centered tightly around 0 meV, typically within ± 100 meV, starting from the self-consistent potential of a converged standard-DFT ground state. The Hubbard parameter for a given site, $I$, is subsequently calculated as the difference in the inverse of two unique response functions.

$$U = (\chi_0^{-1} - \chi^{-1})_{II} \quad (6)$$

The first ($\chi_0$) represents the bare, noninteracting response evaluated after a single iteration of the self-consistent field calculation and the second ($\chi$) is the self-consistent, interacting response evaluated following convergence. Subtraction of the two response functions effectively removes the total-energy curvature associated with the instantaneous rehybridization induced by the applied potential, leaving only the Hubbard $U$ of the interacting system.

At its conception, the methodology described above explicitly neglected higher-order terms in the Coulomb interaction (i.e., those described by Hund's $J$) in favor of a simplified description of the on-site Coulomb repulsion. The resulting Hubbard $U$ was taken to be an effective linear combination of both the true Hubbard $U$ and Hund's $J$.

$$U_{eff} = U - J \tag{7}$$

Although equation 7 is ubiquitous in literature, its form has yet to be rigorously justified.[13] Cococcioni and coworkers later proposed an extension of the linear response method to the calculation of $J$.[12] Notably, they found that the correct insulating ground state of cubic CuO could be reproduced only with the inclusion of Hund's $J$, as the original DFT+$U$ formulation neglected stabilizing magnetic interactions, thus predicting a metallic ground state. These interactions are reintroduced to the functional $E_U$ according to the following expression.

$$E_{\mathrm{U}} = \sum_{I,\sigma} \frac{U^I - J^I}{2} \, \mathrm{Tr}[\mathbf{n}^{I\sigma}(1 - \mathbf{n}^{I\sigma})] + \sum_{I,\sigma} \frac{J^I}{2} \, \left(\mathrm{Tr}[\mathbf{n}^{I\sigma}\mathbf{n}^{I-\sigma}] - 2\delta^{\sigma\sigma_{\min}} n^{I\sigma}\right) \tag{8}$$

The spin index $\sigma_{\min}$ specifically denotes the minority spin channel. The method for calculating $J^I$ closely follows that for the calculation of $U^I$. Here, the response matrix elements take the form

$$\chi_{IJ} = \frac{\partial m_I}{\partial \beta_J} \tag{9}$$

which defines the response of local magnetization ($m_I = n_I^\uparrow - n_I^\downarrow$) to an applied magnetic perturbation, $\beta m_J$. Hund's $J$ is then calculated according to equation 10.

$$J = -(\chi_0^{-1} - \chi^{-1})_{II} \tag{10}$$

Importantly, the functionality of this technique is limited to the calculation of $J$ from a non-magnetic ground state in order to preserve linearity of the response matrices. This limitation is a direct consequence of the fact that the total energy of a system is not variational with respect to its magnetization, as ground state magnetization will typically tend toward saturation.[13]

We employ this extended linear response method to calculate independently the Hubbard $U$ and Hund's $J$ parameters for the Fe 3$d$ subspace of hematite. Starting from a fully relaxed standard-DFT ground state, local potential shifts ($\alpha^I$ and $\beta^I$) ranging from -60 to 60 meV are sequentially applied in 20 meV increments to a single Fe center isolated in a 2×2×2 rhombohedral supercell of hematite. In order to obtain an accurate non-interacting response function, the accuracy threshold of the first iterative Hamiltonian diagonalization of each perturbation was fixed to its final value at convergence of the preceding DFT ground-state calculation. Due to the symmetry-equivalence of the iron centers in hematite, the response function can be computed from perturbing a single iron atom. In order to account for the background response of the lattice, we perform an additional set of perturbations in which equivalent potential shifts are applied to every atom in the supercell *except* the isolated iron. Therefore, each response function takes the form of a 2×2 matrix.

To impose self-consistency on the calculated values of $U$ and $J$, the response functions are iteratively recalculated from a DFT+$U$ and DFT+$J$ ground state, respectively. In this scheme, we apply $U_{in}$ ($J_{in}$) and recalculate an updated value, $U_{out}$ ($J_{out}$). This sequence is repeated until the output value is within 1% of the input value. We note that an iterative recalculation of $J$ from a DFT+$J$ ground state without the inclusion of the Hubbard $U$ parameter may not be generally applicable to all materials. Many systems are driven toward a metallic ground state as the value of the applied $J$ parameter increases (i.e., the band gap narrows with increasing $J$). Consequently, iterating on a DFT+$J$ ground state is rendered impossible if the band gap of the system closes. In such cases, one may choose to first compute a self-consistent $U$ and then iteratively compute $J$ from a fixed DFT+$U$ ground state. In the case of hematite, however, increasing both $U$ and $J$ widens the band gap and preserves an insulating ground state.

*Ab Initio Electron-Phonon Coupling via Vibrational Averages*

The earliest electron-phonon coupling computations were based on perturbative methods aimed toward the direct calculation of electron-phonon matrix elements.[46-48] Here, we use the non-perturbative method of vibrational averages, which involves computing electronic observables within a supercell subjected to a set of atomic displacements that approximate an average thermal configuration, thus circumventing explicit calculation of the interaction matrix. This approach was first developed by Williams[49] and Lax[50] and has recently been extended to the computation of temperature-dependent band structures and optical absorption spectra by Zacharias, et al.[51-53] and Monserrat, et al.[54-56] Notably, perturbative methods of computing electron-phonon matrix elements are limited to include only terms up to a given order. In the vibrational averages approach, this limit is surpassed by virtue of performing full electronic structure calculations on a given configuration. Thus, higher order terms are implicitly captured in all computed observables.[57]

Following the adiabatic Born-Oppenheimer approximation, the temperature-dependent expectation value of an electronic observable can be defined as

$$\left\langle \widehat{O}(T) \right\rangle = \tfrac{1}{\mathcal{Z}} \sum_i \left\langle \Phi_i(\mathbf{u}) \middle| \widehat{O}(\mathbf{u}) \middle| \Phi_i(\mathbf{u}) \right\rangle e^{-E_i/k_B T} \qquad (11)$$

where $|\Phi_i(\mathbf{u})\rangle$ represents nuclear states of energy $E_i$ described by a nuclear configuration expressed in terms of normal mode coordinates, $\mathbf{u} = \{u_{m\mathbf{q}}\}$. Subscripts $m$ and $\mathbf{q}$ represent the phonon band index and wavevector, respectively. $\mathcal{Z}$ is the canonical partition function and $k_B$ is Boltzmann's constant. The central goal of the method of vibrational averages is to construct a nuclear configuration (or set of configurations), $\bar{\mathbf{u}}_T$, that produces a value of the observable equivalent to the vibrational average at a given temperature, $T$.

$$\widehat{O}(\bar{\mathbf{u}}_T) = \left\langle \widehat{O}(T) \right\rangle \qquad (12)$$

Within the quadratic approximation of the harmonic oscillator, equation 12 can be converted into a set of $3(N-1)$ equations that determine $\bar{\mathbf{u}}_T$, where $N$ is the total number of atomic coordinates.

$$u_{m\mathbf{q}}(T) = S_{m\mathbf{q}} \left( \tfrac{1}{2\Omega_{m\mathbf{q}}} [2n_B(\Omega_{m\mathbf{q}}, T) + 1] \right)^{\tfrac{1}{2}} \qquad (13)$$

In equation 13, $S_{m\mathbf{q}}$ takes the form of a sign matrix with elements ± 1 and $n_B(\Omega_{m\mathbf{q}}, T)$ represents the Bose-Einstein occupation number of a phonon with frequency $\Omega_{m\mathbf{q}}$ at temperature $T$.[54-56]

Real-space representations of vibrational displacements are restricted to phonons with real-valued eigenvectors (i.e., those with $\mathbf{q} = 0$). Therefore, the sampling of finite-wavevector phonons requires the construction of a supercell commensurate with the desired $\mathbf{q}$-grid. Herein, we omit the $\mathbf{q}$ subscript, as all phonons that are considered are effectively "folded" into the center of the first Brillouin zone of the chosen supercell. Equation (13) then yields a set of $2^{3(N-1)}$ atomic configurations, where $N$ now denotes the number of atoms in the supercell whose nuclei have been displaced from their equilibrium positions according to equation 14,

$$\Delta\tau_\kappa(T) = \left( \tfrac{\hbar}{M_\kappa} \right)^{\tfrac{1}{2}} \sum_m S_m u_{m,T} \, e_{\kappa m} \qquad (14)$$

where $M_\kappa$ is the mass of atomic species $\kappa$ and $e_{\kappa m}$ are the phonon eigenvectors obtained from diagonalization of the dynamical matrices.[57-60]

In practice, the methodology described above demands careful consideration of the balance between accuracy and efficiency. Dense sampling of the phonon dispersion curve requires the use of large supercells; however, the configuration space defined by equation 14 scales dramatically with system size and is staggeringly vast for even the smallest supercells. In recent years, a number of sophisticated approaches toward overcoming this challenge have been reported. Zacharias, et al. demonstrated an efficient Monte Carlo integration scheme,[51] a one-shot approach to generating optimized sign matrices,[52] and a rigorous algorithm to construct a single temperature-dependent atomic configuration.[53] Monserrat proposed a method of so-called thermal lines, whereby the temperature dependence of electronic observables are computed along a small set of lines in the configuration space (i.e., by keeping the sign matrices constant as temperature is varied).[54] In this work, we adopt an approach that combines Monte Carlo integration with the method of thermal lines in order to generate a small subset of configurations that closely approximate the thermal average.

**Computational Methods**

*Electronic Structure Computations*

First-principles DFT and DFT+$U$+$J$ computations were conducted utilizing the pseudopotential plane wave method implemented in the *Quantum ESPRESSO* package.[61-63] Throughout this work, we employed the exchange-correlation functional proposed by Perdew, Burke, and Ernzerhof and revised for solids (PBEsol)[64, 65] and Optimized Norm-Conserving Vanderbilt (ONCV)[66, 67] pseudopotentials. Converged values of the plane wave cutoff energy and $k$-point grid density were determined for a 10-atom rhombohedral primitive cell of hematite with nuclei clamped at their equilibrium positions. A high plane wave energy cutoff of 1100 eV was chosen to ensure total energy and total interatomic forces were converged to within 10 meV and 0.10 meV/Å per atom, respectively. Self-consistent field (SCF) optimization of the Kohn-Sham wavefunctions was converged on a 4×4×4 Monkhorst-Pack[68] $k$-point grid, sampling 28 symmetry-weighted $k$-points in the first Brillouin zone of the primitive cell. For all SCF calculations involving supercells, the $k$-point grid density was reduced commensurate with the dimensions of the supercell. Iterative atomic relaxations were performed utilizing the Broyden-Fletcher-Goldfarb-Shanno (BFGS)[69-71] algorithm until the total interatomic forces were less than 0.03 meV/Å. Lattice parameters were fixed to an average of experimental values reported for hematite.[72, 73] The antiferromagnetic sublattice of hematite (↑↓↓↑ along the principle axis) was constructed by explicitly treating spin-up and spin-down iron centers as unique atomic species with a fixed anti-parallel spin configuration.

*Vibrational Structure Computations*

Full phonon dispersion curves were calculated with density functional perturbation theory (DFPT) as implemented in the *PHonon* code distributed with the *Quantum Espresso* package.[61-63] Starting from a fully relaxed DFT+$U$+$J$ ground state, dynamical matrices were evaluated on a uniform 3×3×3 grid of $q$-points. Finer sampling was then achieved via Fourier interpolation on a denser **q**-grid.

**Results and Discussion**

*Electronic and Vibrational Properties of the Static DFT+U+J Ground State of Hematite*

We begin by using the linear response method for determining $U$ and $J$ to compute electronic and vibrational structures and spectra of the static ground state of hematite, in which all atoms are clamped in their geometrically relaxed positions. Linear response calculations performed within a 2×2×2 (80-atom) rhombohedral supercell of hematite produced a Hubbard $U$ parameter of 3.120 ± 0.006 eV and a Hund's $J$ parameter of 1.535 ± 0.008 eV. Self-consistency was achieved within five iterations; following which both $U_{out}$ and $J_{out}$ were consistently within 0.1% of their respective preceding values (see Figure S1 in Supplementary Information). We confirmed that these values were converged with respect to the volume of the supercell by performing an identical set of linear response calculations within a 3×2×2 (120-atom) supercell. The output parameters were within 2.5% of those reported above. Therefore, we conclude that a 2×2×2 supercell is sufficient for hosting single-atom perturbations and adequately suppresses spurious interactions of the localized potential with its periodic image.

The electronic band structure and optical dielectric spectrum computed from the static DFT+$U$+$J$ ground state of hematite are shown in Figures 1a-b. We chose to apply a rigid shift of +0.5 eV to all conduction band eigenvalues in order to bring the computed dielectric function into agreement with the measured spectrum. Herein, this shift is applied to all electronic band diagrams, electronic density of states plots, and computed dielectric functions. In order to fulfill the $f$-sum rule governing total oscillator strength, all computed optical spectra are subsequently renormalized by a factor of $(1 - \frac{0.5 \text{ eV}}{\hbar\omega})$.[51, 74] With the applied shift, the computed bandgap is approximately 2.4 eV and is characterized by a direct transition midway between zone-center and the **X**-momentum critical point; however, given the weak dispersion of the conduction band edge, several indirect gaps of similar magnitude exist throughout the Brillouin zone. Consequently, there is little value in definitively classifying the static ground state of hematite as either a direct or indirect semiconductor.

Projection of the DFT+$U$+$J$ Kohn-Sham wavefunctions onto an atomic orbital basis (Figure 1a, right panel) indicates that the uppermost valence bands (> -4 eV) are predominantly of O 2$p$ character with minor Fe 3$d$ hybridization. These mildly dispersive bands become maximally hybridized in a narrow region (~ -5 eV) before giving way to the more ionic Fe 3$d$ orbitals that comprise the less dispersive bands at the bottom of the valence continuum (< -5 eV). Unoccupied Fe 3$d$ orbitals give rise to the minimally dispersive conduction bands and clearly exhibit crystal-field splitting characteristic of high-spin octahedral $Fe^{3+}$ ions. The $t_{2g}$ and $e_g$ conduction edges are separated by approximately 1 eV, in agreement with reported experimental values.[75, 76] In accordance with ligand-field theory, the $e_g$ bands exhibit significantly higher O 2$p$ hybridization than their more ionic $t_{2g}$ counterparts. Higher energy (> 4 eV) conduction bands are comprised of highly dispersive Fe 4$s$ orbitals and contribute negligibly to the density of states, particularly when compared to the sharply peaked density of the Fe 3$d$ conduction bands. As a result, the visible and near-UV absorption spectrum of hematite is dominated by optical transitions from O 2$p$ valence bands to Fe 3$d$ conduction bands.

The single-particle optical dielectric spectrum of the static DFT+$U$+$J$ ground state of hematite was obtained by evaluating the momentum matrix elements coupling each pair of valence and conduction band wavefunctions according to equation 15,[61-63]

$$\epsilon_i(\hbar\omega) = \frac{4\pi}{V\,N_{\mathbf{k}}} \sum_{c,v} \sum_{\mathbf{k}} \frac{|\langle \psi_{\mathbf{k}c}^{\sigma}|\hat{\mathbf{p}}|\psi_{\mathbf{k}v}^{\sigma}\rangle|^2}{(E_{\mathbf{k}c}-E_{\mathbf{k}v})^2} \delta(E_{\mathbf{k}c} - E_{\mathbf{k}v} - \hbar\omega) \tag{15}$$

where $V$ represents the volume of the primitive cell, $N_{\mathbf{k}}$ is the total number of $\mathbf{k}$-points sampled, and $\hat{\mathbf{p}}$ is the momentum operator. Equation 15 does not account for the finite lifetime of the optically excited state; therefore, we chose to apply a homogeneous spectral Gaussian function with a linewidth of 0.3 eV to the computed spectrum in order to approximate lifetime broadening. Herein, this broadening function is applied to all computed spectra. We have previously reported the optical dielectric spectrum of hematite obtained *via* Fresnel analysis of transmission and reflectance spectra measured for a series of polycrystalline thin films of hematite.[40] As shown in Figure 1b, all notable features of the measured dielectric spectrum above 2.4 eV are reproduced in the computed spectrum, with only minor differences in intensity and energetic positions. Importantly, the absorption onset near 2 eV is absent from the computed dielectric function. In the next section, we demonstrate that optical transitions in this region are recovered following the inclusion of electron-phonon coupling.

The DFT+$U$+$J$ phonon dispersion curve of hematite is shown in Figure 1c. As indicated by the projected vibrational density of states, lower-energy phonon modes (< 30 meV) correspond primarily to the displacement of Fe atoms while higher-energy modes (> 50 meV) correspond to displacements of the O atoms. The majority of the optical phonon branches exhibit minimal dispersion and produce strong peaks in the density of states. Most notably, the two highest-energy branches are nearly dispersionless, giving rise to an anomalously high density of states at approximately 81 meV. Notably, two phonon band gaps (i.e., regions of near-zero vibrational density) appear in the phonon dispersion, indicated by the horizonal arrows in Figure 1c. The first is a narrow gap that follows the sharp drop in the density of states just below 50 meV. A second gap appears just below 57 meV and is effectively much wider, as the region between 57 and 62 meV is crossed only by a single, highly dispersive band near zone center. Gaps in the phonon density of states are known to inhibit the decay of high-energy optical phonons into lower-energy optical and acoustic phonons.[77] Pathways for phonon relaxation are subject to energy and momentum conservation; therefore, viable routes for the decay of vibrational population above a gap may be severely limited, thus leading to a buildup of long-lived optical phonons. This phenomenon can profoundly impact the thermalization of hot carriers. If not permitted to decay, high-energy optical phonons emitted during thermal relaxation of an excited carrier can, in turn, prolong the lifetime of the carriers if reabsorbed. As such, we predict that phonon population near these gaps will exhibit strong coupling to photoexcited carriers.

The shaded regions of the dispersion curve indicate branches of predominantly longitudinal character: red and blue regions correspond to longitudinal optical (LO) and longitudinal acoustic (LA) branches, respectively. Details of this characterization can be found in the Supplementary Information. Interestingly, the LA branch exhibits numerous avoided crossings with lower-energy (20–30 meV) LO branches, indicating a high degree of LA/LO phonon hybridization away from zone center. Acoustic/optical hybridization introduces additional carrier-phonon coupling to the acoustic branches and, similar to the occurrence of a phonon band gap, removes viable pathways for the decay of optical phonons into acoustic phonons.[78]

A comparison of the previously measured Raman spectrum of a polycrystalline thin film of hematite under bandedge excitation (2.21 eV) with the DFT+$U$+$J$ vibrational density of states is shown in Figure 1d.[40] Vertical lines indicate the positions of all computed $\Gamma$-point Raman-active phonons. The computed highest-energy mode is positioned at approximately 81 meV; therefore, we attribute all higher-energy (> 90 meV) features of the measured spectrum to second-order, multi-phonon scattering processes.[40] We compare the first-order region of the Raman spectrum (< 90 meV) to the zone-center density of Raman-active LO and transverse optical (TO) modes, obtained by sampling a dense **q**-grid tightly centered around the $\Gamma$-point ($|\mathbf{q}| \leq 2.5 \times 10^4$ cm$^{-1}$). Each of the bands observed in the measured spectrum are accurately reproduced in the computed density spectrum. Notably, the measured bands at 31 and 51 meV correspond to regions of predominately LO-phonon density in the computed $\Gamma$-point spectrum.

Although first-order Raman scattering is restricted to phonons of near-zero wavevector, second-order scattering processes can access the entire phonon density of states while still conserving wavevector. In the simplest cases, these processes involve the simultaneous scattering of two phonons of equal and opposite wavevector from the same branch. In Figure 1d, the second-order region of the measured Raman spectrum is overlaid with the 2-LO and 2-TO density of states. Both density spectra reproduce the second-order Raman features; however, the 2-LO spectrum exhibits the same intensity distribution as that of the measured spectrum. Additionally, the lowest-energy band of the second-order spectrum at approximately 100 meV coincides with twice the computed phonon bandgap at 50 meV, suggesting that two-phonon Raman scattering involves only modes above this gap. Overall, the computed phonon density of states accurately accounts for all of the features observed in the first- and second-order regions of the Raman spectrum. In the following sections, we demonstrate that the resonantly-enhanced second-order

Raman spectrum corresponds to thermally activated optical transitions into polaronic states near the band edge of hematite.

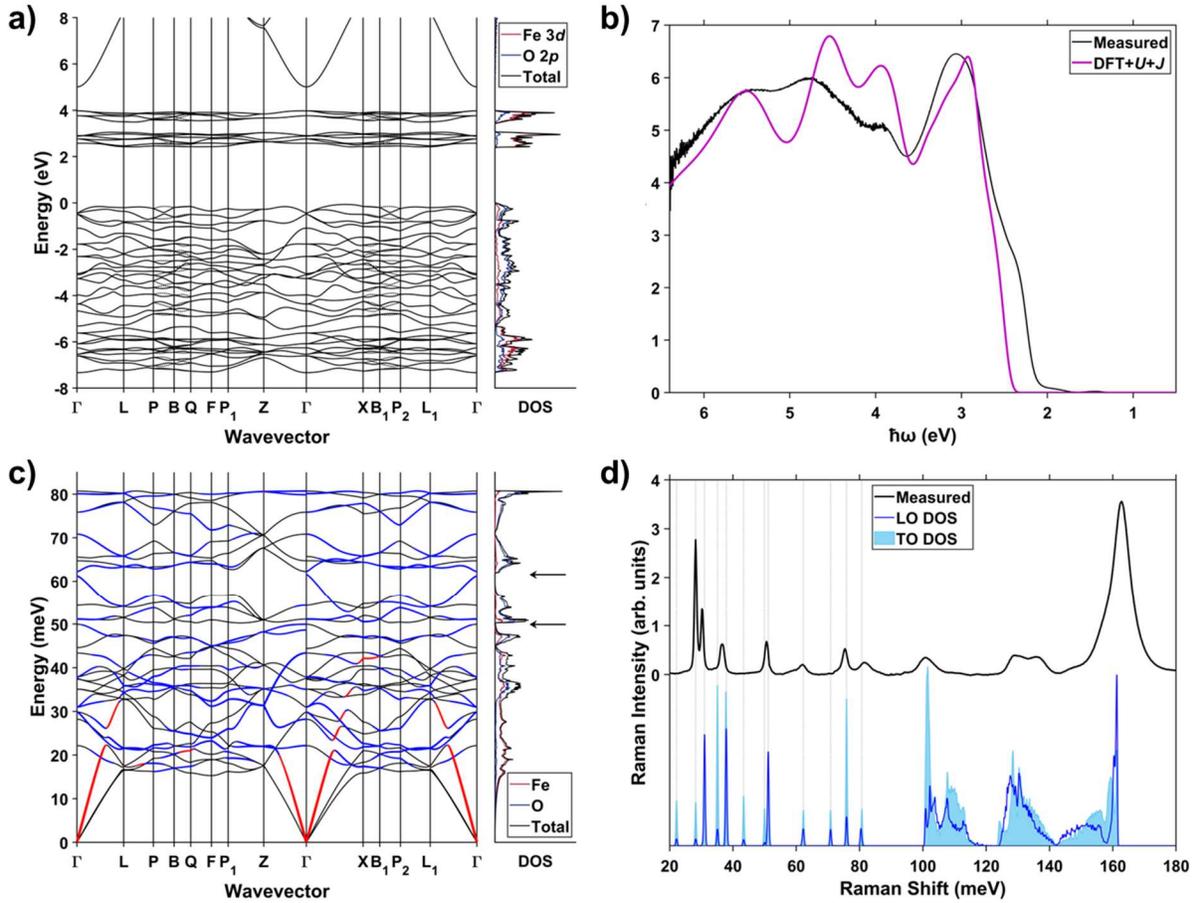

**Figure 1: a)** Electronic dispersion curve (left) and projected density of states (right) computed from the static DFT+$U$+$J$ ground state of hematite. Dashed lines in the dispersion curve represent bands of opposite spin along paths where spin degeneracy is lifted. **b)** Optical dielectric spectrum of hematite obtained from a Fresnel analysis of measured transmission and reflection spectra of a polycrystalline thin film (black line, reproduced from reference 40) and computed single-particle optical dielectric spectrum of the DFT+$U$+$J$ ground state of hematite (purple line). The computed spectrum is artificially broadened with a Gaussian width of 0.3 eV. **c)** Phonon dispersion curve (left) and projected vibrational density of states (right) computed from the static DFT+$U$+$J$ ground state of hematite. LO and LA branches are indicated by blue and red shading, respectively. Horizontal arrows in the density of states indicate phonon band gaps. **d)** Resonance Raman spectrum of a polycrystalline thin film of hematite measured with an excitation energy of 2.21 eV (top, black line, reproduced from reference 40) and computed vibrational density of LO (blue) and TO (light blue) phonons. The first-order region of the computed spectrum (< 90 meV) corresponds to the $\Gamma$-point phonon density of states, while the second-order region (> 90 meV) corresponds to the two-phonon density of states across the entire phonon dispersion curve. Note, the computed spectrum does not account for the Raman cross-section of each band. Dashed vertical lines correspond to the central positions of computed zone-center Raman-active modes.

*Thermal Difference Spectra of Hematite Computed via Ab Initio Electron-Phonon Coupling*

With the self-consistent values of *U* and *J* and the vibrational structure of hematite in hand, we are now positioned to model the temperature dependent electronic structure of hematite using the method of thermal lines combined with Monte Carlo averaging. The experimental data used to benchmark these calculations are a series of temperature dependent thermal difference spectra reported in our previous work.[40] Here, we define a thermal difference spectrum, $\Delta\epsilon_i(\hbar\omega, T)$, to be the difference between the optical dielectric spectrum measured at a particular temperature and that measured at room temperature.

$$\Delta\epsilon_i(\hbar\omega, T) = \epsilon_i(\hbar\omega, T) - \epsilon_i(\hbar\omega, 294\ K) \tag{16}$$

Stochastic Monte Carlo averaging of the computed thermal difference spectrum (TDS) of a 2×2×2 supercell of hematite at 80 K converged rapidly with respect to the number of thermal lines sampled. As shown in Figures 2a-b, the result obtained by averaging spectra computed across a random distribution of 25 thermal lines is nearly identical to that obtained by sampling only six lines. In fact, we demonstrate that even a *single* thermal line can sufficiently reproduce the Monte Carlo average (Figure 2c). A 2×2×2 supercell of hematite permits real-space representation of 237 phonon modes within the primitive Brillouin zone, specifically those at the **Z**, **F**, **L**, **L₁**, and **Γ** critical points. In Figure 2d, we demonstrate that stochastic sampling of the configuration space of a larger, 3×2×2 supercell (357 phonon modes), produces a similar average TDS at 80 K. We therefore conclude that a 2×2×2 supercell is sufficiently large for computing temperature-dependent optical spectra of hematite. Finally, we note that all the computed TDS shown in Figure 2 accurately reproduce the previously reported measured spectrum at 80 K[40] and recover the band-edge optical transitions absent from the computed static dielectric spectrum (Figure 2 insets).

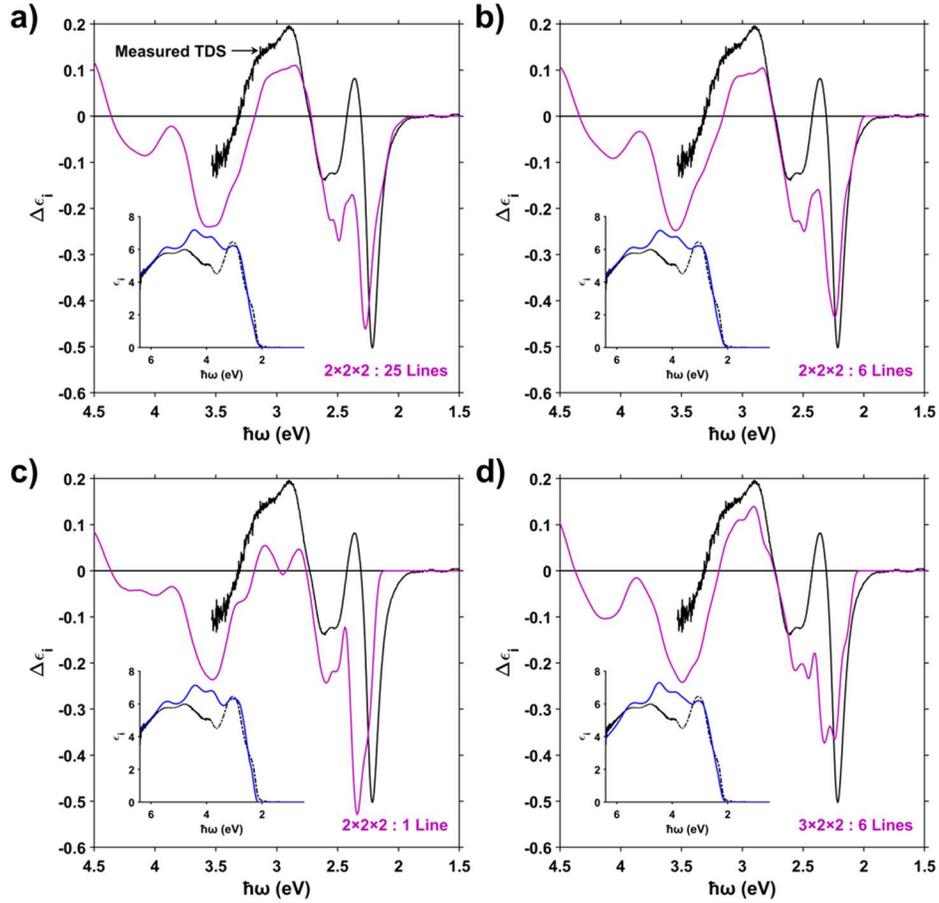

**Figure 2:** Computed TDS of hematite at 80 K (purple lines) obtained from Monte Carlo averaging of a random distribution of **a)** 25 thermal lines and **b)** 6 thermal lines within a 2×2×2 supercell. TDS at 80 K computed from **c)** a single thermal line within a 2×2×2 supercell that closely approximates the Monte Carlo average and **d)** a random distribution of 6 thermal lines within a 3×2×2 supercell. For comparison, the measured TDS at 80 K of a polycrystalline thin film of hematite is shown in all panels (solid black lines). The insets of each figure illustrate comparisons of the computed room-temperature (294 K) optical dielectric spectra (blue lines) with that measured for a thin film of hematite (dashed black lines).

We previously reported TDS of a hematite thin film measured from 30 to 573 K.[40] Here, we compute temperature-dependent TDS averaged across five thermal lines of a 2×2×2 supercell of hematite, each closely approximating the Monte Carlo average. Several representative comparisons of the computed and measured spectra are shown in Figures 3a-d. At low temperatures (< 294 K), the positions and relative intensities of the features in the measured TDS are accurately reproduced in the computed spectra (Figures 3a-b). We note that, due to the UV absorption edge of the cryostat windows, these measurements were restricted to photon energies below 3.5 eV. High-temperature (> 294 K) spectra were collected without the use of a cryostat and, therefore, extend to 6.4 eV. In Figure 3c we compare the measured and computed TDS at 473 K, illustrating that the computed spectrum reproduces the near-UV features of the measured TDS, albeit with a significant deviation in the band-edge intensity.

We characterized the temperature-dependence of the computed and measured spectra by integrating the absolute value of $\Delta\epsilon_i$ over the interval 1.0–3.5 eV (Figure 3d). We define the total intensity to be the negative absolute value for spectra collected at temperatures below 294 K and the positive absolute value for spectra collected at temperatures above 294 K. The measured and computed values display remarkable agreement, with minor deviations arising at temperatures above 400 K. Both sets of data were fit to a Bose-Einstein distribution function according to equation 17.

$$\int \Delta\epsilon_i(\hbar\omega, T) = A \times \Delta n_B(\hbar\Omega, T) \qquad (17)$$

Here, $A$ is a constant that corresponds to the value of $\int \Delta\epsilon_i$ at 0 K and $\Delta n_B$ represents the change in the Bose-Einstein occupation number of a phonon of energy $\hbar\Omega$ relative to its occupation at room temperature. The temperature dependence of the computed TDS fits the thermal distribution of a 50.3 ± 0.1 meV phonon, while the measured spectra correspond to that of a 49.2 ± 1.6 meV phonon. These values suggest that the temperature-dependent growth of optical transitions near the band gap of hematite coincides with population of phonons above a 50-meV threshold.

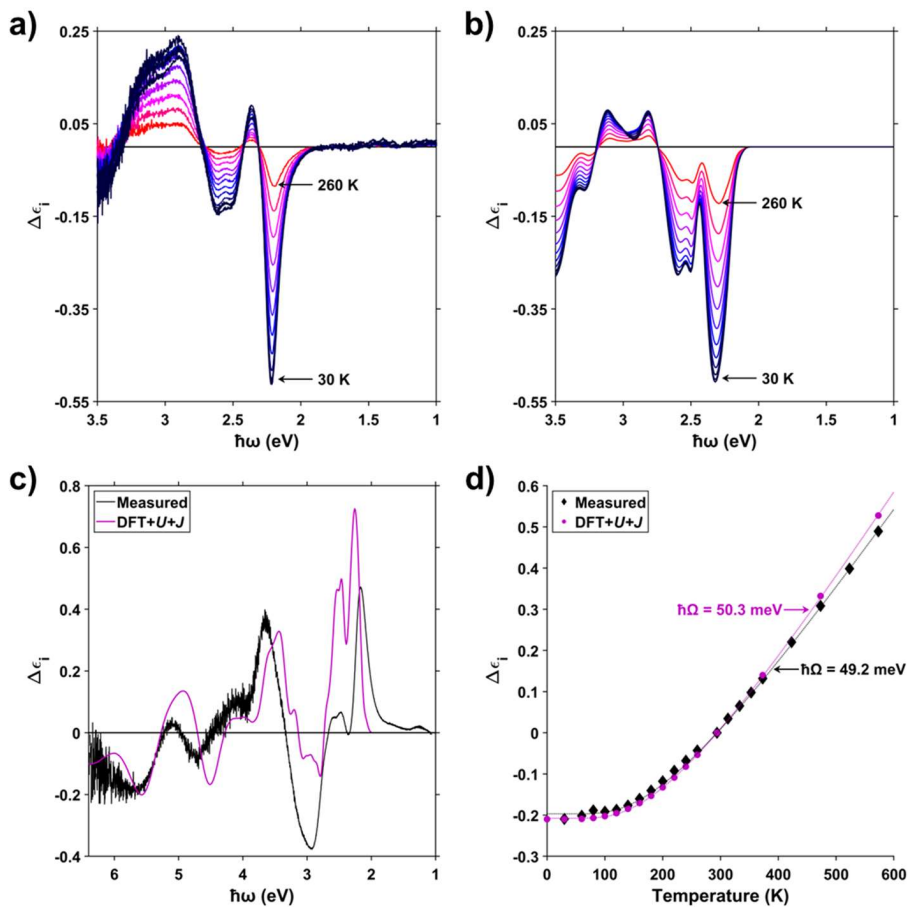

**Figure 3: a)** Measured and **b)** computed TDS of hematite below room temperature (T < 294 K). Computed spectra represent the average of five thermal lines within the configuration space of a 2×2×2 supercell of hematite that closely approximate the Monte Carlo average at 80 K. Part **a)** is reproduced from reference 40, copyright 2021 American Chemical Society. **c)** A comparison of the computed (purple line) and measured (black line) TDS of hematite at 473 K. **d)** The integrated intensity of the computed (purple circles) and measured (black diamonds) TDS across the investigated temperature range. Dashed lines indicate the Bose-Einstein distribution function fit to the computed (purple line) and measured (black line) values.

*Correspondence Between Thermal Difference and Resonance Raman Spectra of Hematite*

In this section, we use the method of thermal lines to show that including displacements along only the eigenvectors corresponding to the phonon modes that are most strongly coupled to the near band-edge excitation of hematite, as evidenced by resonance Raman excitation profiles, can reproduce all of the features observed in the thermal difference spectra. Our previously reported resonance Raman spectra of a hematite thin film collected with excitation photon energies spanning the visible range are shown in Figure 4a.[40] Relative scattering intensities of the observed bands exhibit a strong dependence on the excitation photon energy. Most notably, the second-order spectrum (> 90 meV) is observed only for excitations that are resonant with band-edge transitions (2.0–2.5 eV). In Figure 4b, we show the excitation profile of the integrated intensity of each second-order Raman band. The bands at 100, 129, and 136 meV share a similar excitation profile and are strongly enhanced at 2.21 and 2.53 eV. Importantly, each of these profiles closely trace the band-edge TDS features. The 163-meV band is similarly enhanced at 2.21 eV; however, it exhibits markedly less enhancement at 2.53 eV. This discrepancy is addressed in the following section. These results strongly suggest that the TDS and second-order Raman spectrum of hematite arise from the same phenomenon.

The majority of the first-order modes (< 90 meV) in the observed Raman spectrum share a similar excitation profile (see Figure S2 in Supplementary Information), with the exception of the three shown in Figure 4c: the 31, 51, and 81-meV bands. Both the 31 and 51-meV bands correspond to regions of high LO-phonon density computed at zone center (see Figure 1d). Similar to the second-order modes, the 31 and 51-meV bands exhibit notable enhancement at photon excitation energies of 2.21 and 2.53 eV, respectively. The excitation profile of the 81-meV phonon is nearly identical to that of the 100, 129, and 136-meV bands and also closely reproduces the relative intensity of the band-edge TDS. We note that this band may be, in part, a two-phonon band arising from the combination of the 31 and 51-meV bands. Complete assignment of all conceivable combinations is beyond the scope of this work. Instead, we draw the general conclusion that the second-order Raman spectrum and band-edge features of the TDS arise from optical transitions coupled to phonon population above the 50-meV phonon band gap.

We further support this conclusion by performing a series of calculations similar to those used above to generate the computed thermal difference spectra, but with the inclusion of only phonons of energy greater than 50 meV where the spatial phases of these phonons have been synchronized such that their displacements are localized to a single iron center. We find that atomic displacements along the coordinates of the LO phonon modes above the 50-meV gap are sufficient to reproduce the features of the thermal difference spectrum (Figure 4d). This result, combined with the correspondence of the 2-LO density of states with the intensity distribution of the second-order Raman spectrum (see Figure 1d), indicates that the optical transitions near the band gap exhibit strong coupling to LO phonons with energies greater than 50 meV.

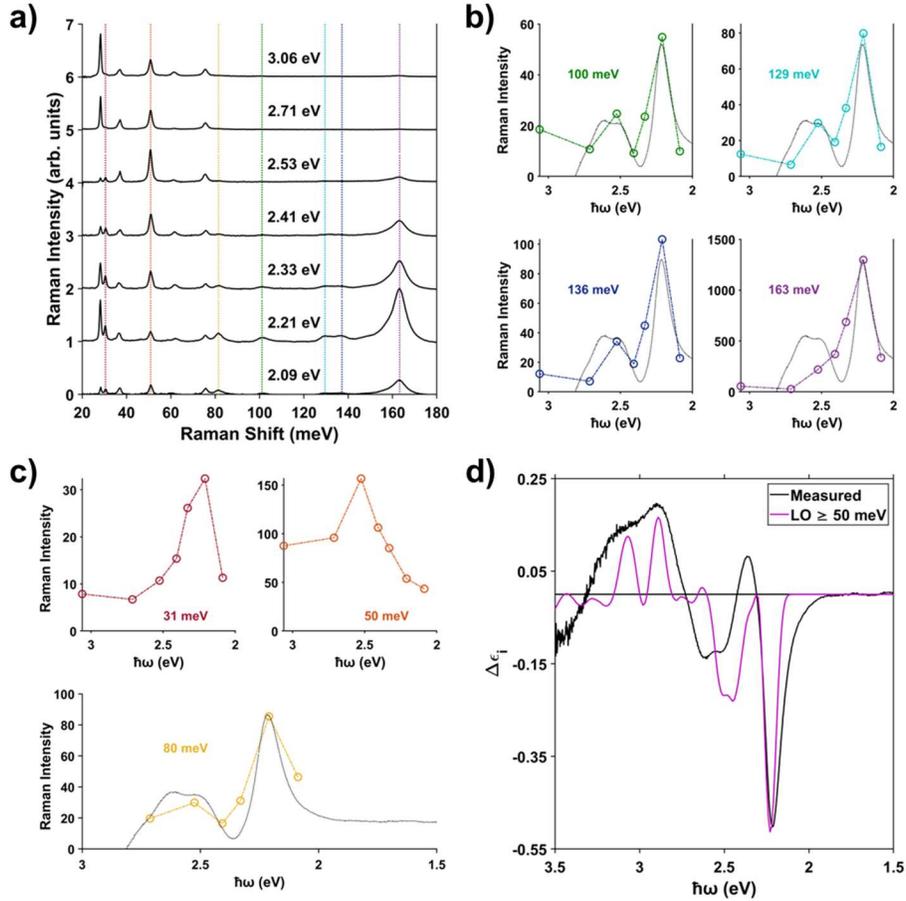

**Figure 4: a)** Resonance Raman spectra measured from a polycrystalline hematite thin film using seven visible excitation photon energies. Data reproduced from reference 40, copyright 2021 American Chemical Society. Excitation profiles are shown for the bands indicated by dashed vertical lines. These include **b)** all second-order modes and **c)** first-order modes that exhibit a unique excitation profile. Intensities represent the integral of the full band width. All second-order profiles and the profile of the 81-meV band are overlaid with the TDS measured at 80 K. For comparison, we have inverted the measured TDS such that $\Delta\epsilon_i = \epsilon_i(294\ K) - \epsilon_i(80\ K)$. **d)** TDS measured at 80 K (black line) compared to that computed with displacements along only the coordinates of LO phonons with energies above the 50-meV phonon band gap.

*Thermally Activated Polaronic Optical Transitions in Hematite*

Longitudinal optical (LO) phonons are known to play a critical role in the formation of polarons due to the strong lattice polarizations induced by their propagation.[79] In this section, we show that these polarizations result in the formation of a series of polaronic band tails that give rise to the features of the TDS. We connect our model of the temperature-dependent electronic structure of hematite to polaron formation by identifying bands in the electronic dispersion curves that appear at finite temperature and evaluating the localization of wavefunctions associated with these bands.

To assign the transitions that appear in the optical dielectric spectrum at finite temperature, we compute electronic dispersion curves at two temperatures (294 and 573 K) in the configuration space of a 2×2×2 supercell of hematite along a single thermal line, that which best approximated the Monte Carlo average of the TDS at 80 K (see Figure 2c). The supercell band structures were unfolded to span the first Brillouin zone of the primitive cell according to the method proposed by Zacharias and Giustino.[53] The resulting dispersion curves, along with their corresponding densities of states, are shown in Figures 5a-b. The bands are represented by a spectral function, $\rho$, that effectively represents a momentum-resolved density of states. Numerous phonon sidebands appear in the dispersion curve at finite temperature, illustrating the profound strength of electron-phonon coupling in hematite, particularly within the Fe $3d$ orbitals. Two minimally dispersive band tails appear just below both the $t_{2g}$ and $e_g$ conduction bands, with wavefunctions strongly localized to two neighboring $Fe^{3+}$ ions. Additionally, a weakly dispersive tail appears at the valence band edge, moderately localized to the $O^{2-}$ ions surrounding the two $Fe^{3+}$ ions. The features of the computed TDS arise primarily from optical transitions coupling the uppermost valence bands to the localized conduction band tails.

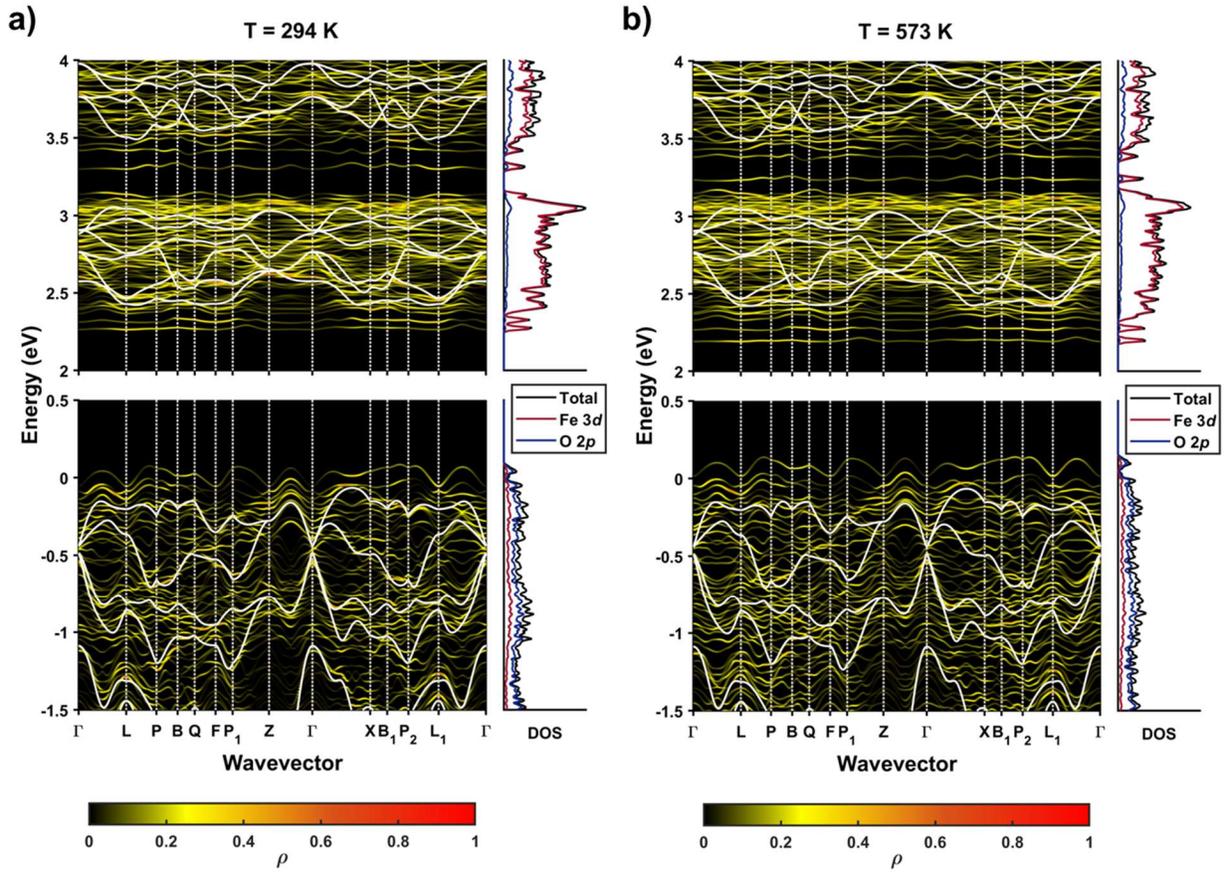

**Figure 5:** Finite-temperature electronic dispersion curves of a 2×2×2 supercell of hematite at **a)** 294 K and **b)** 573 K along a single thermal line. The dispersion curves have been unfolded to span the space of the first Brillouin zone of a primitive rhombohedral cell of hematite and are each represented by a three-dimensional spectral function, $\rho$, which corresponds to the momentum-resolved density of states. Projected densities of states are shown in the panels on the right-hand side of each band dispersion diagram. Bands computed from the static DFT+$U$+$J$ ground state are represented by solid white lines.

As seen in the temperature-dependent densites of states plotted in Figure 6a, the $t_{2g}$ and $e_g$ band tails become less dispersive and more localized with increasing temperature, and exhibit a significantly stronger thermal shift than their corresponding continua. Importantly, these conduction band tails are present at 0 K and therefore dictate the zero-point band gap renormalization. Thus, the features of the TDS do not indicate the *formation* of new optical transitions with increasing temperature, but rather the *evolution* of transitions that are always present. At the zero point, the $t_{2g}$ band tails are seperated by 30 meV. This splitting increases monotonically with temperature, with the lower energy band shifting at a faster rate than its higher-energy counterpart. The $e_g$ band tails exhibit a similar behavior, but are initially split by a larger magnitude (83 meV). As such, both pairs of band tails give rise to similar features in the observed TDS, with the $e_g$ absorption band notably wider than the $t_{2g}$ band (Figure 6b). At all finite temperatures, the valence band tail and uppermost $t_{2g}$ conduction band tail are separated from their respective continuum edges by approximately 80 meV. We therefore propose the unique excitation profile of the 163-meV second-order Raman band (Figure 4b) is the result of an additional double-resonance enhancement at near band-edge excitation.

Real-space depictions of the charge density ($\sum_{\mathbf{k}}|\psi_{\mathbf{k}c}|^2$) of the conduction band tails are shown in Figure 6c. Both pairs are strongly localized within the volume of a 2×2×2 supercell, with the $e_g$ wavefunctions marginally more disperse due to hybridization with surrounding O $2p$ orbitals. The localization arises from a significant phonon-induced distortion of the octahedral ligand fields around isolated iron centers. Notably, the charge densities of the two $t_{2g}$ band tail states are localized around two adjacent pairs of Fe atoms, indicating that these states may provide a pathway for transport of localized charge via a thermally activated hopping mechanism.

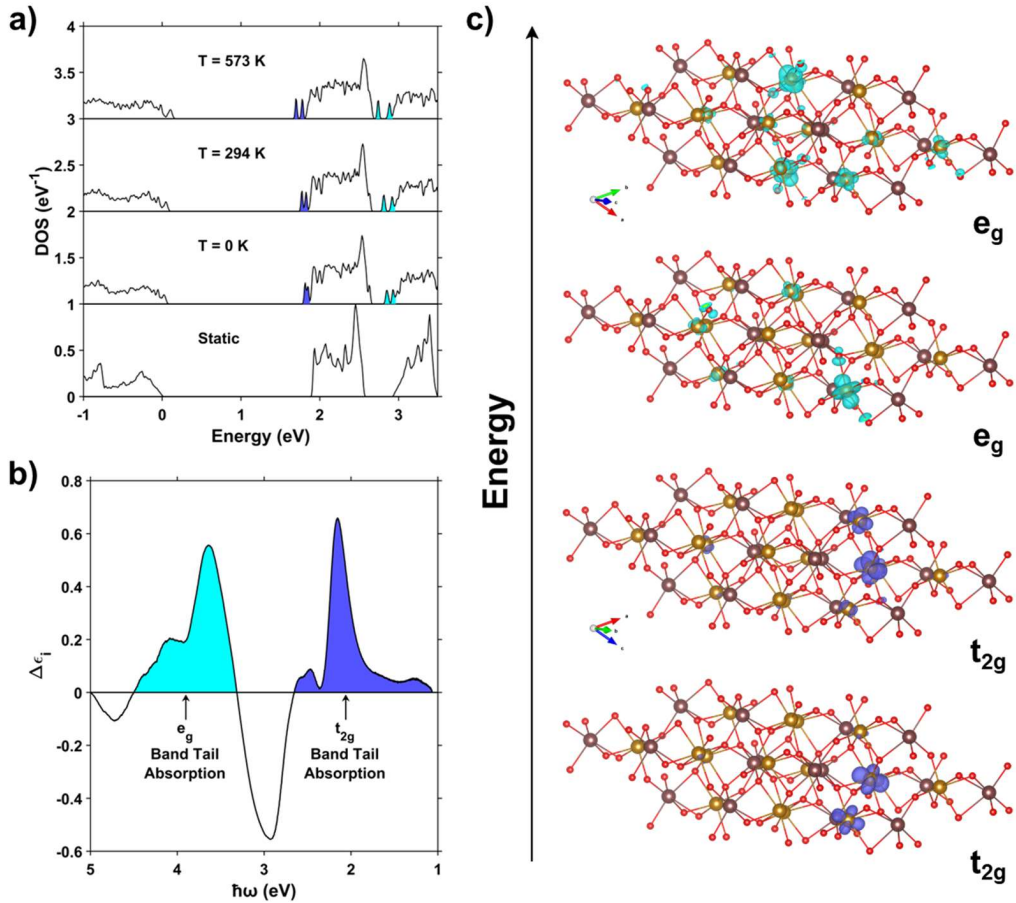

**Figure 6: a)** Evolution of the electronic density of states within a 2×2×2 supercell of hematite with increasing temperature. The bottom panel shows the density of states computed from a static primitive cell of hematite, with nuclei clamped at equilibrium positions. The shaded regions indicate density associated with the $t_{2g}$ (purple) and $e_g$ (light blue) conduction band tails. **B)** Assignment of the features observed in the TDS measured at 573 K. Purple and light blue shaded regions indicate the absorption features associated with optical transitions from the valence band edge to the $t_{2g}$ and $e_g$ conduction band tails, respectively. **c)** Real-space representations of the charge density associated with the $t_{2g}$ (purple) and $e_g$ (light blue) conduction band tail wavefunctions at 294 K. Isosurfaces represent the average probability density of the wavefunctions computed across eight equally weighted **k**-points. Charge densities are ordered from top to bottom with decreasing energy. $Fe^{3+}$ ions are represented by yellow and brown spheres, distinguishing spin-up and spin-down centers, and $O^{2-}$ ions are represented by red spheres.

Finally, we demonstrate that population of the t$_{2g}$ conduction band tail directly forms an electron small polaron. An excess electron introduced to a thermally distorted 2×2×2 supercell at 294 K fully localizes to the lowest-energy t$_{2g}$ band tail. Following geometrical relaxation of the singly charged state, all atoms but those in the vicinity of the localized charge return to their pristine equilibrium positions. The polaron charge density is shown in Figure 7a and is nearly identical to the charge density of the vacant band tail, with the amplitude more equally distributed across the two Fe centers. Additionally, the charge density is fully contained within the volume of a single primitive cell, indicating the formation of an electron small polaron. Conversely, an identical charge introduced to a pristine 2×2×2 supercell, with nuclei clamped at their equilibrium positions, delocalizes across the entire lattice (see Figure 7a). Therefore, we conclude that a thermal population of phonons is capable of inducing localized lattice polarizations that serve as nucleation sites for small polaron formation.

The displacement vectors associated with the relaxed polaron are shown in the inset of Figure 7c. The distortion extends over two octahedral Fe sites and is characterized by the compression of the Fe-Fe distance and the symmetric displacement of oxygen ions from the center of the distortion. We express the contribution of a particular phonon to the polaronic distortion ($C_{m\mathbf{q}}$) according to equation 18, where $\Delta\tau_p$ is the atomic displacement associated with the polaron and $d_{pm\mathbf{q}}$ are normalized phonon displacement vectors.

$$C_{m\mathbf{q}} = \langle \Delta\tau_p | d_{pm\mathbf{q}} \rangle \tag{18}$$

Here, the subscript $p$ specifically denotes atoms within the polaronic distortion, namely two iron atoms and ten oxygen atoms. Figure 7b contains a plot of the phonon dispersion curve of hematite shaded to indicate the phonon branches that contribute most significantly to the polaronic distortion ($C_{m\mathbf{q}} > 0.5$). For each phonon band $m$, Figure 7c plots the sum of $C_{m\mathbf{q}}$ over all $\mathbf{q}$ points as a function of phonon energy. As demonstrated in Figures 7b-c, the strongest contributions to the distortion arise from 81-meV phonons at the **Γ** and **Z**-momentum critical points. These phonons account for displacements of the oxygen atoms in the distortion. The iron displacements are comprised of multiple low-energy optical phonons, the most prominent of which is the zone-center 31-meV phonon. Additional contributions arise primarily from the hybridized acoustic and optical branches (20–30 meV) at the bottom of the dispersion curve.

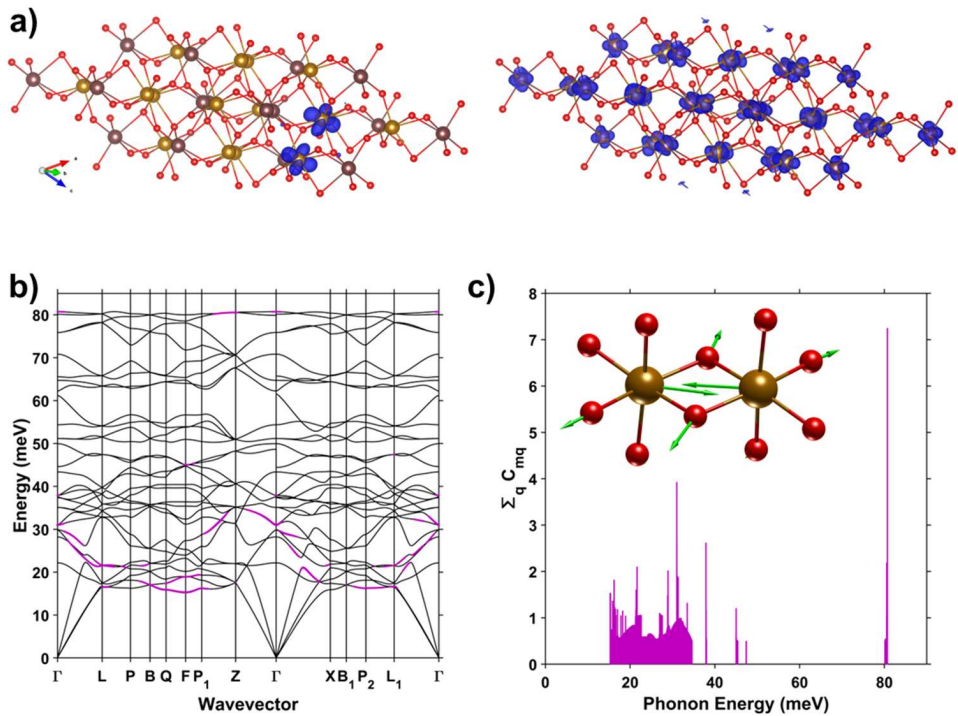

**Figure 7: a)** Geometrically relaxed charge density of an electron introduced to a thermally distorted 2×2×2 supercell of hematite at 294 K (left) and to a pristine lattice (right). The isosurfaces represent only the contribution of the lowest occupied band. Charge densities were computed across eight equally weighted **k**-points. **b)** Phonon dispersion curve of hematite indicating the branches that contribute to the displacement vectors of the polaronic distortion (purple). **c)** Spectral representation of the coefficient expansion of the polaron displacement vectors. The spectrum is effectively the phonon density of states that strongly contribute to the polaronic distortion. The inset shows the displacement vectors associated with the relaxed polaronic distortion. The vectors (green arrows) are computed by subtracting the atomic coordinates of the pristine lattice from those of the polaron-distorted lattice.

**Conclusion**

We have demonstrated that, when combined, DFT+$U$+$J$ and the method of vibrational averages can accurately reproduce the effects of electron-phonon coupling in hematite. Our results indicate that a thermal distribution of phonons gives rise to highly-localized bandtails at the edges of the $t_{2g}$ and $e_g$ conduction bands. We attribute these bands to strong lattice polarizations arising from population of the LO phonons above the 50-meV phonon bandgap. Following an electronic excitation, population of these bandtails leads directly to the formation of an electron small polaron. The resulting distortion lies along primarily the coordinates of the 81 and 31-meV optical phonons. Notably, this distortion encompasses two adjacent Fe atoms, which implicates a potential pathway for electron polaron hopping, a crucial mechanism for photoconductivity in this material.

This work reperesents a first-principles theoretical description of carrier-phonon coupling in hematite that supports the conclusion of our previous work: thermally activated optical transitions directly populate electron small polaron states in hematite. By computing an atomically precise decription of carrier-phonon coupling, we are able to unambiguously determine which phonons contribute to the polaronic distortion. These assignments are confirmed by experimental resonance Raman and thermal difference spectra first reported in our previous work.[40] We anticipate that, when extended to other transition metal oxides, this computational approach will reveal fundamental insights into mechanisms of polaron formation in optically excited states of these matierials.

**Supplementary Material**

Details of the convergence of the *U* and *J* parameters calculated from the linear response method, characterization of the phonon modes as LO, LA, TO or TA, and additional resonance Raman profiles of first-order modes are provided in the supplementary material.

**Acknowledgements**

Financial support for this work was provided by the National Science Foundation under grant number CHE-2044462. The authors are grateful for the computational resources and technical support provided by the University of Rochester Center for Integrated Research Computing.

**Author Declarations**

**Conflict of Interest**

The authors have no conflicts to declare

**Author Contributions**

**Jacob L. Shelton:** Conceptualization (equal); Formal analysis (lead); Investigation (lead); Methodology (lead); Software (lead); Visualization (lead); Writing - original draft (lead); Writing - review & editing (equal). **Kathryn E. Knowles:** Conceptualization (equal); Methodology (supporting); Visualization (supporting); Writing - review & editing (equal); Project administration (lead); Supervision (lead); Funding acquisition (lead)

**Data Availability**

The data that support the findings of this study are available from the corresponding author upon reasonable request.

# Supporting Information for

# Polaronic Optical Transitions in Hematite (α-Fe₂O₃) Revealed by First-Principles Electron-Phonon Coupling

Jacob L. Shelton and Kathryn E. Knowles*

Department of Chemistry, University of Rochester, Rochester, NY 14627

*Corresponding Author e-mail: kknowles@ur.rochester.edu

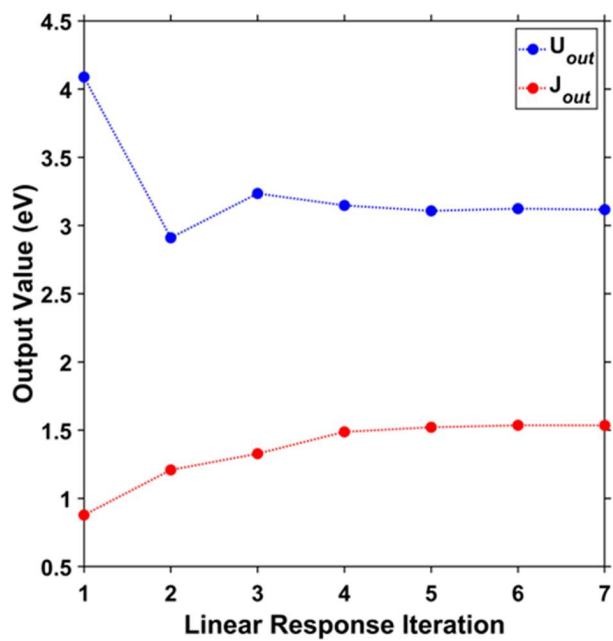

**Figure S1:** Self-consistent convergence of $U_{out}$ (blue circles) and $J_{out}$ (red circles) obtained from linear response calculations performed within a 2×2×2 (80-atom) rhombohedral supercell of hematite. Within five iterations, both $U_{out}$ and $J_{out}$ were consistently within 0.1% of their respective preceding values.

**Characterization of Phonon Modes**

Phonon modes are characterized as longitudinal optical (LO), longitudinal acoustic (LA), transverse optical (TO), and transverse acoustic (TA) according to the following equations.

$$c_{m\mathbf{q}}^{LO} = \sum_{\kappa,\kappa'} \langle \mathbf{q}|d_{\kappa m\mathbf{q}}\rangle \times (1 - \langle d_{\kappa' m\mathbf{q}}|d_{\kappa m\mathbf{q}}\rangle)$$

$$c_{m\mathbf{q}}^{LA} = \sum_{\kappa,\kappa'} \langle \mathbf{q}|d_{\kappa m\mathbf{q}}\rangle \times \langle d_{\kappa' m\mathbf{q}}|d_{\kappa m\mathbf{q}}\rangle$$

$$c_{m\mathbf{q}}^{TO} = \sum_{\kappa,\kappa'} (1 - \langle \mathbf{q}|d_{\kappa m\mathbf{q}}\rangle) \times (1 - \langle d_{\kappa' m\mathbf{q}}|d_{\kappa m\mathbf{q}}\rangle)$$

$$c_{m\mathbf{q}}^{TA} = \sum_{\kappa,\kappa'} (1 - \langle \mathbf{q}|d_{\kappa m\mathbf{q}}\rangle) \times \langle d_{\kappa' m\mathbf{q}}|d_{\kappa m\mathbf{q}}\rangle$$

Here, $d_{\kappa m\mathbf{q}}$ are the normalized displacement vectors of a particular phonon mode $m$ of wavevector $\mathbf{q}$. Atom indices are represented by $\kappa$. The phonon displacement vectors are related to the phonon eigenvectors such that $|d_{\kappa m\mathbf{q}}\rangle = M_\kappa^{-1/2}|e_{\kappa m\mathbf{q}}\rangle$. The term $\langle \mathbf{q}|d_{\kappa m\mathbf{q}}\rangle$ represents the inner product of the atomic displacements with the propagation direction of the phonon. As such, the magnitude of $\langle \mathbf{q}|d_{\kappa m\mathbf{q}}\rangle$ is directly proportional to the longitudinal character of the mode. The term $\langle d_{\kappa' m\mathbf{q}}|d_{\kappa m\mathbf{q}}\rangle$ represents the inner product of the displacement vector of atom $\kappa$ with that of all adjacent atoms $\kappa'$. As such, the magnitude of $\langle d_{\kappa' m\mathbf{q}}|d_{\kappa m\mathbf{q}}\rangle$ directly proportional to the acoustic character of the mode. All vectors are normalized such that the maximum value of both $\langle \mathbf{q}|d_{\kappa m\mathbf{q}}\rangle$ and $\langle d_{\kappa' m\mathbf{q}}|d_{\kappa m\mathbf{q}}\rangle$ is unity. Thus, $1 - \langle \mathbf{q}|d_{\kappa m\mathbf{q}}\rangle$ and $1 - \langle d_{\kappa' m\mathbf{q}}|d_{\kappa m\mathbf{q}}\rangle$ represent the transverse and optical character of the mode, respectively.

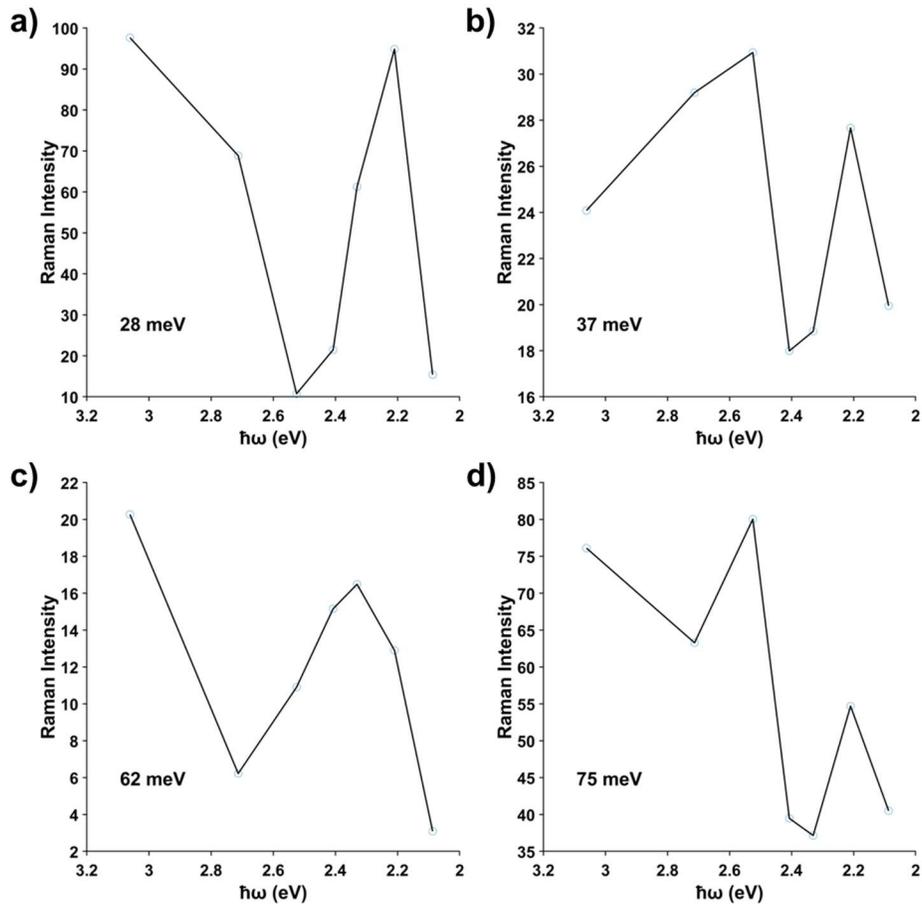

**Figure S2:** Resonance Raman excitation profiles for the first-order **a)** 28-meV, **b)** 37-meV, **c)** 62-meV, and **d)** 75-meV modes of the measured Raman spectra of a polycrystalline thin film of hematite. Each mode displays a similar excitation profile with significant enhancement in a narrow range near 2.21 eV, followed by broader enhancement at excitation energies above 2.6 eV.